\documentclass[14pt,a4paper]{article}

\usepackage[english]{babel}
\usepackage{amsmath,amssymb,amsfonts,amsthm}
\usepackage[mathscr]{eucal}
\usepackage[all]{xy}
\usepackage{hyperref}
\usepackage{setspace}
\usepackage{upgreek}
\usepackage{hyperref}

\usepackage{times}
\usepackage{color}

\usepackage{indentfirst}

\begin{document}

\title{\textbf{Conservation laws and stability of higher derivative\\ extended Chern-Simons}}
\author {V.~A.~Abakumova\footnote{abakumova@phys.tsu.ru}, \ D.~S.~Kaparulin\footnote{dsc@phys.tsu.ru}, \ and S.~L.~Lyakhovich\footnote{sll@phys.tsu.ru}}
\date{\footnotesize\textit{Physics Faculty, Tomsk State University, Tomsk 634050, Russia }}

\maketitle

\begin{abstract}
\noindent The higher derivative field theories are notorious for the stability problems both at classical and quantum level. Classical instability is connected with unboundedness of the canonical energy, while the unbounded energy spectrum leads to the quantum instability. For a wide class of higher derivative theories, including the extended Chern-Simons, other bounded conserved quantities which provide the stability can exist. The most general gauge invariant extended Chern-Simons theory of arbitrary finite order $n$ admits $(n - 1)$-parameter series of conserved energy-momentum tensors. If the $00$-component of the most general representative of this series is bounded, the theory is stable. The stability condition requires from the free extended Chern-Simons theory to describe the unitary reducible representation of the Poincar\'e group. The unstable theory corresponds to nonunitary representation.
\end{abstract}

\section{Introduction}

Theories with higher derivatives are important in modern theoretical
physics. The most known examples of such models are the
Pais-Uhlenbeck (PU) oscillator \cite{PU}, generalized Podolsky
electrodynamics \cite{PPRD42}, modified theories of gravity
\cite{BLCQG87,SFPRL10,FTLRR10,LPPRL11}, and conformal higher spin
theories \cite{FTPR85}. Theories with higher derivatives have higher
symmetry and better convergency properties at the classical and
quantum level compared to their counterparts without higher
derivatives, but also they are known for classical and quantum
instability.

Stability is an important characteristic of dynamics. Classical
stability means that the motions of theory are finite at every
moment of time. For mechanical models it can be provided by
existence of the Lyapunov function. For theories without higher
derivatives the canonical energy plays the role of Lyapunov
function. So, if the canonical energy is bounded, the theory is
classically stable. Quantum stability means that quantum system has
a well-defined vacuum state with the lowest possible energy. This
type of stability is provided by the bounded Hamiltonian being the
phase-space equivalent of the canonical energy.

Let us consider the stability problem of higher derivative theories.
The theory with second time derivatives, described by the action
functional
\begin{equation}\label{}
\displaystyle S[x^i(t)]=\int dt\, L(x^i,\dot{x}^i,\ddot{x}^i)\,,
\end{equation}
have the following canonical energy:
\begin{equation}\label{}
\displaystyle E^{can}=\ddot{x}^i\frac{\partial
L}{\partial\ddot{x}{}^{i}}+\dot{x}^i\Big(\frac{\partial L}{\partial
\dot{x}^i}-\frac{d}{dt}\frac{\partial L}{\partial
\ddot{x}^i}\Big)-L\,.
\end{equation}
This expression is linear in $\dddot{x}^i$\,, and so it is
unbounded. The unbounded canonical energy cannot serve as a Lyapunov
function of the system, and it cannot ensure classical stability of
the model. The canonical Hamiltonian is also unbounded. So, the
model is not stable at quantum level.

In \cite{KLShEPJC14}, it was shown that a wide class of theories
with higher derivatives admits, except for canonical energy, another
conserved quantities. If some of these conserved quantities are
bounded, they can ensure classical and quantum stability of the
model. First such conservation laws were introduced for the
Pais-Uhlenbeck oscillator \cite{BKAPP05, DSJPA06}. Later the
stability of higher derivative theories was studied in
\cite{SS09,PPLA13, AVVJMP17,KKLEPJC15}. We will apply this idea to
study the stability problem for the higher derivative extended
Chern-Simons theory \cite{DJPLB99}.

The rest of the paper is organized as follows. In Section 2, we
consider the stability of the Pais-Uhlenbeck oscillator \cite{PU}
from the viewpoint of existence of conservation law, which is
different from the canonical energy. In Section 3, we analyze the
stability condition for the extended Chern-Simons of arbitrary
finite order. In Section 4, the general construction is exemplified
by the third-order extension of Chern-Simons theory. In Conclusion
we summarize the results.

\section{Toy model}

In this section, we illustrate how the idea of stabilizing dynamics
by the conserved quantities works in the simplest higher derivative
theory.

The PU oscillator of fourth order is a theory of a single dynamical
variable $x(t)$ with the action functional
\begin{equation}\label{PU4}
\displaystyle S[x(t)]=\int L(x,\dot{x},\ddot{x})dt\,,\qquad
L(x,\dot{x},\ddot{x})=\frac{1}{2}\big(-\ddot{x}^2+(\omega_1^2+\omega_2^2)\dot{x}^2-\omega_1^2\omega_2^2x^2\big)\,.
\end{equation}
Here, the parameters $\omega_i,\, i=1,2,$ are the frequencies of
oscillations. We assume that the frequencies are different and
nonzero,
\begin{equation}\label{}
\phantom{\frac12}\omega_1\neq\omega_2\,,\qquad
\omega_1^2+\omega_2^2\neq 0\,.\phantom{\frac12}
\end{equation}
The Euler-Lagrange equation for the model (\ref{PU4}) has the form,
\begin{equation}\label{}
\displaystyle \phantom{\frac12}\frac{\delta S}{\delta x}\equiv x^{(4)}+(\omega_1^2+\omega_2^2)\ddot{x}+\omega_1^2\omega_2^2x^2=0\,.\phantom{\frac12}
\end{equation}
The solution to this equation is the bi-harmonic oscillation,
\begin{equation}\label{xtPU}
\displaystyle
\phantom{\frac12}x(t)=A_1\sin(\omega_1t+\varphi_1)+A_2\sin(\omega_2t+\varphi_2)\,,\phantom{\frac12}
\end{equation}
where the amplitudes $A_i$, and initial phases $\varphi_i,\, i=1,2,$
are integration constants. The motion is finite,
\begin{equation}\label{}
\displaystyle \phantom{\frac12}|x(t)|\leq
|A_1|+|A_2|\,.\phantom{\frac12}
\end{equation}
Thus, the Pais-Uhlenbeck oscillator is a stable model.

Let us explain the stability of the PU theory from the viewpoint of
the conserved quantities of the model. There are two-parameter
series of symmetries of the action functional (\ref{PU4}),
\begin{equation}\label{Sym-PU4}
    \phantom{\frac12}\delta_\beta x=\beta_1
    \dot{x}+\beta_2\dddot{x}\,,\phantom{\frac12}
\end{equation}
where $\beta_1,\beta_2$ are infinitesimal transformation parameters,
being constants. The first symmetry in the set is time translation.
The second transformation in (\ref{Sym-PU4}) is the higher symmetry.
The two-parameter series of conserved quantities is associated with
these symmetries by the Noether theorem:
\begin{equation}\label{}
    \phantom{\frac12}J=\beta_1E^{can}+\beta_2 J^2\,,\phantom{\frac12}
\end{equation}
where $E^{can}$ is the canonical energy of the model, and $J^2$ is
another independent integral of motion,
\begin{equation}\label{EPU}\begin{array}{rl}\displaystyle
E^{can}&\displaystyle=\Big(\dot{x}x^{(3)}-\frac{1}{2}\ddot{x}^2\Big)+
\frac{1}{2}\Big((\omega_1^2+\omega_2^2)\dot{x}^2+\omega_1^2\omega_2^2x^2\Big)\,;
\\[5mm]\displaystyle
J^2&\displaystyle=\frac{1}{2}\Big(\dddot{x}^2+(\omega_1^2+\omega_2^2)\ddot{x}^2\Big)+
\omega_1^2\omega_2^2\Big(x\ddot{x}-\frac{1}{2}\dot{x}^2\Big)\,.
\end{array}\end{equation}
Both of these quantities are unbounded quadratic forms of initial data
$\dot{x},\ddot{x},\dddot{x},\ddddot{x}$, but they can be joined in
two bounded combinations,
\begin{equation}\label{JiPU}\begin{array}{c}
    \phantom{\frac12}J_1=J^2+\omega_1^2 E^{can}\,,\qquad J_2=J^2+\omega_2^2
    E^{can}\,,\phantom{\frac12}\\[5mm]\displaystyle
    \phantom{\frac12}J_i=\frac{1}{2}\Big((x^{(3)}+\omega_i^2\dot{x})^2+(\omega_1^2+\omega_2^2-\omega_i^2)(\ddot{x}+\omega_i^2x)^2\Big)\,,
    \qquad i=1,2\,.\phantom{\frac12}
\end{array}\end{equation}
Any bounded combination of these bounded quantities with positive
coefficients is a positive-definite quadratic form of initial data.

The positive definite conserved quantity, being constructed from the
integrals of motion (\ref{JiPU}), selects the stationary bounded
surface in the phase-space of the theory. This bounded conserved
quantity stabilizes the dynamics of the PU theory.

\section{Extended Chern-Simons}

Consider $3d$ Minkowski space with the local coordinates
$x^\mu,\,\mu=0,1,2\,,$ and the metric
\begin{equation}\label{}
\displaystyle
\phantom{\frac12}\eta_{\mu\nu}=\text{diag}(+1,-1,-1)\,.\phantom{\frac12}
\end{equation}
The extended Chern-Simons is a gauge theory of vector field $A=A_\mu(x) dx^\mu$ with the action functional
\begin{equation}\label{SChS}
\displaystyle
\phantom{\frac12}S[A(x)]=\frac{m^2}{2}\sum_{p=1}^{n}\bigg(\alpha_p\int
A_\mu(x) F^{(p)\mu}(x)d^3x\bigg)\,.\phantom{\frac12}
\end{equation}
Here, $m$ is a parameter with dimension of mass, and dimensionless
real constants $\alpha_1,\ldots,\alpha_n$ are parameters of the
model. Without loss of generality we assume that $\alpha_n\neq0$, and the
notation is used:
\begin{equation}\label{FChS}
\displaystyle
F^{(p)}{}{}_{\mu}=m^{-p}\varepsilon_{\mu\nu\rho}\partial^\nu
F^{(p-1)\rho}\,,\qquad F^{(0)}{}_{\mu}\equiv A_\mu\,,\qquad \qquad
r=1,\ldots,n\,,
\end{equation}
where $\varepsilon$ denotes the $3d$ Levi-Civita symbol with $\varepsilon_{012}=1$.
The Euler-Lagrange equations for the action functional (\ref{SChS}) have the form
\begin{equation}\label{LeqChS}
\displaystyle \frac{\delta S}{\delta A}\equiv
m^2\sum_{p=1}^{n}\alpha_pF^{(p)}=0\,.
\end{equation}
These equations involve the $n$-th time derivatives of $A$\,.

The action (\ref{SChS}) is Poincar\'e-invariant. The space-time
translations are symmetries of the action functional (\ref{SChS}),
\begin{equation}\label{PUsim}
\displaystyle \phantom{\frac12}\delta_\xi A_\mu=\xi^\nu\partial_\nu
A_\mu\,,\phantom{\frac12}
\end{equation}
where $\xi$ is the transformation parameter. The canonical
energy-momentum is associated with this symmetry,
\begin{equation}\label{Tmunucan}
\displaystyle T^{can}_{\phantom{(}\mu\nu}(\alpha)=
\frac{m^2}{2}\sum_{p=1}^{n}\sum_{r+s=p}
\alpha_p\Big(F^{(r)}{}_{\mu}F^{(s)}{}_{\nu}+
F^{(r)}{}_{\nu}F^{(s)}{}_{\mu}-
\eta_{\mu\nu}\eta^{\rho\sigma}F^{(r)}{}_{\rho}F^{(s)}{}_{\sigma}\Big)\,.
\end{equation}
The $00$-component of the energy-momentum tensor has the form
\begin{equation}\label{Tmunucan}
\displaystyle T^{can}_{\phantom{(}00}(\alpha)=
\frac{m^2}{2}\sum_{\mu=0}^{2}\sum_{p=1}^{n}\sum_{r+s=p} \alpha_p
F^{(r)}{}_{\mu}F^{(s)}{}_{\mu}\,.
\end{equation}
This quantity is linear in $F^{(n-1)}$ for $n>2$,
\begin{equation}\label{}
\displaystyle T^{can}_{\phantom{(}00}(\alpha)=
\frac{m^2}{2}\alpha_nF^{(n-1)}{}_{0} F^{(1)}{}_{0}+\ldots\,,
\end{equation}
the dots denote the terms that do not include $F^{(n-1)}$. Thus, the
energy in the extended Chern-Simons theory is unbounded whenever the
higher derivatives are included in the Lagrangian.

The series of higher symmetries, which generalize (\ref{PUsim}) of
the action functional (\ref{SChS}) has the form
\begin{equation}\label{}
\displaystyle \delta_\xi
A_\mu=\sum\limits_{q=1}^{n-1}\beta_q\xi^\nu\partial_\nu
F^{(q-1)}{}_{\mu}\,,
\end{equation}
where $\beta_q,\, q=1,\ldots, n-1,$ are the parameters of symmetry
series, being real constants. The $(n-1)$-parameter series of
conserved tensors is connected with this symmetry by the Noether
theorem,
\begin{equation}\label{TmunuChS}
\displaystyle T_{\mu\nu}(\alpha,\beta)=
\sum\limits_{r,s=1}^{n-1}C_{r,s}(\alpha,\beta) \Big(F^{(r)}{}_{\mu}
F^{(s)}{}_{\nu}+F^{(r)}{}_{\nu}
F^{(s)}{}_{\mu}-\eta_{\mu\nu}\eta^{\rho\sigma}
F^{(r)}{}_{\rho}F^{(s)}{}_{\sigma}\Big)\,.
\end{equation}
The Bezout matrix $C_{r,s}(\alpha,\beta)$ of two polynomials is defined by the generating relation
\begin{equation}\label{CChS}
\displaystyle C_{r,s}(\alpha,\beta)=\frac{\partial^{r+s}}{\partial^rz\,\partial^sz'}\Big(\frac{M(z)N(z')-M(z')N(z)}{z-z'}\Big)\Big|_{z=z'=0}\,,
\end{equation}
where  $z$ and $z'$ are two independent variables, and
\begin{equation}\label{MNChS}
\displaystyle M(z)=\sum\limits_{p=1}^{n}\alpha_pz^p\,, \qquad N(z)=\sum\limits_{q=1}^{n-1}\beta_qz^{q}\,.
\end{equation}
The representatives of the series (\ref{TmunuChS}) are defined by the formula
\begin{equation}\label{}
\displaystyle T^{(q)}_{\phantom{(}\mu\nu}(\alpha)=\frac{\partial T_{\mu\nu}(\alpha,\beta)}{\partial \beta_q}\,, \qquad q=1,\ldots,n-1\,.
\end{equation}
By construction, $T^{(1)}_{\phantom{(}\mu\nu}\equiv T^{can}_{\phantom{(}\mu\nu}$, and other conserved tensors are independent.

The $00$-component of general conserved tensor (\ref{TmunuChS}) has
the form
\begin{equation}\label{T00ChS}
\displaystyle T_{00}(\alpha,\beta)=\frac{m^2}{2}\sum_{\mu=0}^2
\sum\limits_{r,s=1}^{n-1}C_{r,s}(\alpha,\beta)F^{(r)}{}_{\mu}F^{(s)}{}_{\mu}\,.
\end{equation}
This quantity is a quadratic form of the variables
$F^{(r)}{}_{\mu}$. So, it is bounded if
\begin{equation}\label{stcond}
\displaystyle \phantom{\frac12}C_{r,s}(\alpha,\beta)\quad \text{is a
positive definite matrix}\,.\phantom{\frac12}
\end{equation}
This condition is a restriction on the parameters $\beta$ in the
series of energy-momentum tensors (\ref{TmunuChS}). It is
consistent, iff the polynomial
\begin{equation}\label{}
\displaystyle M'(z)=\sum\limits_{q=0}^{n-1}\alpha_{q+1}z^q
\end{equation}
has simple and real roots. From the viewpoint of the representation
theory, it means that the stability condition requires from the free
extended Chern-Simons theory to describe the unitary reducible
representation of the Poincar\'e group. If the roots of $M'(z)$ are
multiple or complex, there is no bounded integral of motion that can
stabilize the dynamics.

\section{Extended Chern-Simons of order 3}

Let us demonstrate the general construction in the case $n=3$. The
action functional of the model reads
\begin{equation}\label{LChS3}
\displaystyle
S[A(x)]=\frac{1}{2}\int\Big(\alpha_3m^{-1}\varepsilon_{\mu\nu\rho}\partial^\nu
G^\rho+\alpha_2G_\mu+\alpha_1mF_\mu\Big)\,, \quad \alpha_3\neq0\,,
\end{equation}
where the notation is used
\begin{equation}\label{}
\displaystyle\phantom{\frac12} G_\mu\equiv
F^{(2)}{}_\mu=m^{-2}\big(\partial_\mu\partial^\nu
A_\nu-\partial_\nu\partial^\nu A_\mu\big)\,, \quad F_\mu\equiv
F^{(1)}{}_\mu= m^{-1}\varepsilon_{\mu\rho\nu}\partial^\rho
A^\nu\,.\phantom{\frac12}
\end{equation}
The model (\ref{LChS3}) is invariant under the following
two-parameter series of symmetries:
\begin{equation}\label{}
\displaystyle \phantom{\frac12}\delta_\xi
A_\mu=\beta_1\xi^\nu\partial_\nu A_\mu+\beta_2\xi^\nu\partial_\nu
F_\mu\,,\phantom{\frac12}
\end{equation}
The corresponding two-parameter series of conserved tensors reads
\begin{equation}\label{TChS3}
\displaystyle T_{\mu\nu}(\alpha,\beta)=\beta_1T^{can}_{\phantom{(}\mu\nu}(\alpha)+\beta_2T^{(2)}_{\phantom{(}\mu\nu}(\alpha)\,,
\end{equation}
where
\begin{equation}\label{}
\displaystyle
\phantom{\frac12}T^{can}_{\phantom{(}\mu\nu}(\alpha)=\frac{m^2}{2}\Big(\alpha_3\big(G_\mu
F_\nu+G_\nu F_\mu-\eta_{\mu\nu}G_\rho
F^\rho\big)+\alpha_2\big(2F_\mu F_\nu-\eta_{\mu\nu}F_\rho
F^\rho\big)\Big)\phantom{\frac12}
\end{equation}
is the canonical energy-momentum, and
\begin{equation}\label{}
\displaystyle
\phantom{\frac12}T^{(2)}_{\phantom{(}\mu\nu}(\alpha)=m^2\Big(\alpha_3\big(G_\mu
G_\nu-\frac{1}{2}\eta_{\mu\nu}G_\rho G^\rho\big)-\alpha_1\big(F_\mu
F_\nu-\frac{1}{2}\eta_{\mu\nu}F_\rho
F^\rho\big)\Big)\phantom{\frac12}
\end{equation}
is another independent conserved tensor. The canonical energy is
linear in $G_\mu$ and unbounded,
\begin{equation}\label{}
\displaystyle T^{can}_{\phantom{(}00}(\alpha)=m^2\sum\limits_{\mu=0}^2\Big(\alpha_3G_\mu F_\mu+\frac{1}{2}\alpha_2F_\mu F_\mu\Big)\,.
\end{equation}
The $00$-component of the general conserved tensor (\ref{TChS3}) has
the form
\begin{equation}\label{}
\displaystyle T_{00}(\alpha,\beta)=\frac{m^2}{2}\sum\limits_{\mu=0}^2\Big(\beta_2\alpha_3G_\mu G_\mu+2\beta_1\alpha_3G_\mu F_\mu+(\beta_1\alpha_2-\beta_2\alpha_1)F_\mu F_\mu\Big)\,.
\end{equation}
It can be bounded. The boundedness condition for the quadratic form reads
\begin{equation}\label{}
\displaystyle \phantom{\frac12}\alpha_3\beta_2>0\,, \quad
-\alpha_3\beta_1^2+\alpha_2\beta_1\beta_2-\alpha_1\beta_2^2>0\,.\phantom{\frac12}
\end{equation}
It is consistent if parameters of the model (\ref{LChS3}) satisfy the condition
\begin{equation}\label{}
\displaystyle
\phantom{\frac12}\alpha_2^2-4\alpha_1\alpha_3>0\,.\phantom{\frac12}
\end{equation}
According to the representation theory, the stable theory corresponds to one of the two cases: theory of two self-dual massive spin 1 with different masses, or theory of massless spin 1 and massive spin 1 subject to a self-duality condition \cite{TPNPLB84, DJPLB84}.

\section{Conclusion}

We considered a class of vector field models whose wave operator is a polynomial in the Chern-Simons operator. We demonstrated that the gauge theory of order $n$ admits $(n-1)$-parameter series of conserved tensors, whose $00$-component can be bounded, while the canonical energy is always unbounded for $n>2$. The bounded conservation laws ensure the stability of dynamics at classical level. At quantum level the stability is provided by bounded Hamiltonian. The constrained Hamiltonian formulations with bounded Hamiltonian of the extended Chern-Simons were constructed in \cite{AKLEPJC18,AKLRPJ18}.

\vspace{0.2cm} \noindent
{\bf Acknowledgments.} This research was funded by the state task of Ministry of Science and Higher Education of Russian Federation, grant number 3.9594.2017/8.9.

\end{document}